# On-chip label-free plasmonic based imaging microscopy for microfluidics

P. Arora[1, 2, *] and A. Krishnan[1, 2]

[1]Centre for NEMS and Nanophotonics (CNNP), [2]Experimental Optics Laboratory (Expo)
Department of Electrical Engineering, Indian Institute of Technology Madras, Chennai - 600036, India
*E-mail address: pankaj.arora@mail.huji.ac.il

**Abstract:**

In this work, we demonstrated on-chip label-free imaging microscopy using real and Fourier Plane (FP) microscopic dark field images of Surface Plasmons (SP), excited on engineered 1D and 2D low aspect ratio periodic plasmonic nanostructures. The periodic plasmonic nanostructures with period ≈ resonance wavelength, were engineered to exhibit transmission resonances as transmission peaks in visible spectrum at normal incidence, without using extraordinary optical transmission phenomena. The plasmonic nanostructures exhibited a polarization rotation of $90°$ mediated by differential phase retardation in the SP mode due to Transverse Electric and Magnetic components. This was used to develop a dark field on-chip plasmonic polarization microscope for imaging SP excitation in real and Fourier planes. After successful integration of these plasmonic nanostructures with SU- 8 based microfluidic channels, a real-time monitoring of label-free on-chip sensing was demonstrated. Label-free on-chip imaging for interface of colorless miscible and immiscible analytes flowing on plasmonic nanostructures in the microfluidic channels were performed using color-selective filtering nature of plasmonic nanostructures. Since the imaging is realized on a chip and does not need any complicated and bulky arrangement, it will be well suited for on-chip point of care diagnostics.

**Introduction:**

Traditional labels such as fluorophores or chromophores are widely used in microfluidic applications to visualize flow or detect the presence or concentration of relevant species[1–4]. However, each of these labels has shortcomings including bleaching of fluorophores and non-specificity of chromophores etc. [5,6]. In particular in microfluidics, labeling with multiple dyes may suffer from the low Reynolds number laminar flow and when introduced at high concentration, may alter the physical properties of the flow solution [7,8]. To avoid this, label-free on-chip sensing has gained huge attention because of low sample consumption, avoiding time consuming labeling process and the efficient delivery of target analytes to sensing sites [9–12]. In particular, Surface Plasmon Resonance (SPR) based label-free optical technique has been integrated with microfluidics to measure the changes in the refractive index near the surface of the sensor with the advantage of simple collinear broadband illumination and portable spectrophotometer[13–16]. Moreover, periodic nanostructure coupled SPRs have been proven to be a promising tool in label free imaging due to the color-selective filtering nature and the possibility of coupling to Surface Plasmons (SP) modes at normal incidences [17–21]. In this work, we demonstrate an on-chip label free plasmonic based imaging microscopy for microfluidics on engineered periodic nanostructures operating in collinear transmission mode. The engineered fabricated 1D and 2D nanostructures after successful integration of SU-8 based microfluidic channels were placed in between two crossed polarizers ($\theta_P$ = 45° and $\theta_A$ = 135°) to diminish direct $0^{th}$ order transmission [22] and capture bright SPs emission against a dark background in real and

Fourier Plane (FP) images. The analytes of interest with different refractive indices were introduced and the change in the color corresponding to SP excitation wavelengths was captured in real and FP images and quantified with CMYK components of images extracted using image processing [23]. The interaction of two colorless miscible and immiscible liquid analytes were captured by real and FP images without any stain (label) in SU-8 based microfluidic channels using color-selective filtering nature of plasmonic nanostructures. The most common configuration for hydrodynamic focusing is a 3-inlets-1-outlet device which was fabricated and tested that allowed to capture the flow of air bubbles on plasmonic nanostructures with real and FP images through hydrodynamic focusing. Since the imaging is realized on a chip and does not need any complicated and bulky arrangement, it will be well suited for on-chip point of care diagnostics. Several of these proposed structures can be integrated on the same chip with plasmonic nanostructures of different sizes and shapes to realize multiplex, multianalyte sensing for several different target analytes.

**Fabrication of plasmonic on-chip device:**

1D and 2D periodic plasmonic nanostructures with period ≈ resonance wavelength were designed using Rigorous coupled wave analysis to exhibit transmission resonances as transmission peaks instead of dips in visible spectrum at normal incidence [24]. We have recently shown that, by sandwiching a thin layer of homogeneous metal between the patterned metal and glass substrate, it is possible to convert the signature of SP from transmission dips to transmission peaks [22]. Thin sandwiched layer between patterned metal and the substrate leads to leakage radiation at specific angles corresponding to SP mode. The engineered nanostructures were fabricated on glass substrate coated with a thin gold film of thickness 30 nm, using e-beam lithography, gold evaporation of thickness 30 nm and lift-off process [17,22]. Inclusion of a homogeneous 30 nm metal layer below the periodic structures resulted in mitigating the charging effects during electron beam lithography. Finally, on-chip plasmonic device was fabricated by integrating the fabricated plasmonic substrates with microfluidic channels. Figure 1(a) shows the schematic of the on-chip device integrated with plasmonic substrate and Figure 1(b) shows the SEM micrograph for fabricated 1D and 2D plasmonic nanostructures.

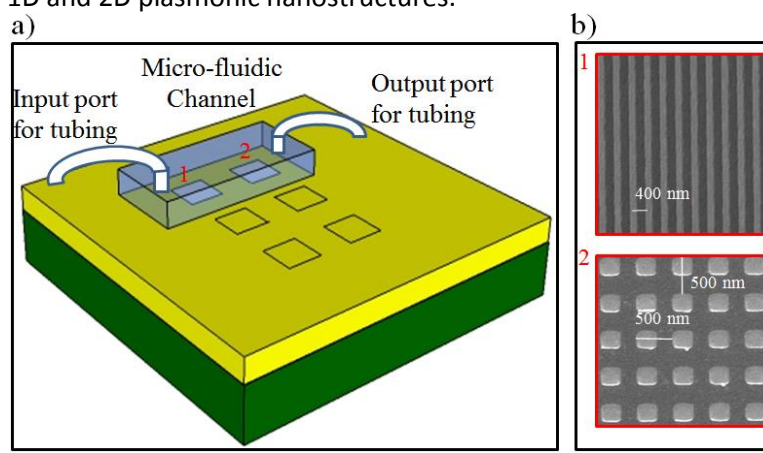

Fig. 1(a) : The schematic for on-chip plasmonic device integrated with microfluidic channel (b) SEM micrograph for fabricated periodic plasmonic 1D and 2D nanostructure

After fabricating the plasmonic substrates, SU-8 (10) resist from Microchem was spin coated on the surface of plasmonic substrates at the speed of 5000 rpm for 30 sec. to a thickness of 10 μm. It was then soft baked at 80 °C for 20 minutes. The required channel mask designed in Clewin software and printed on transparency sheet, was used to expose the SU-8 layer using UV- photolithography. The alignment between fabricated plasmonic nanostructures and microfluidic channels was carried out using mask aligning process. The substrates were prebaked at 80°C for 15 minutes and then patterns were developed in SU-8 developer for 1 minute, followed by rinsing with DI water. Again the baking step was done by keeping the samples in oven at 80°C temperature for 30 minutes. Now, the next step was to cap and seal the fabricated SU-8 microfluidic channels. A PDMS cap was used with holes for inlets and outlets corresponding to fabricated channels and aligned with SU-8 reservoirs. The property of epoxy group of SU-8 to form covalent bond with amine group was used to seal PDMS with SU-8 channels. The surface of PDMS was first functionalized with (−OH) groups through oxygen plasma for 30 sec. exposure and then was immersed in a solution of (3-AminoPropyl) Tri Ethoxy Silane (APTES) for 2 hours and as a result of which, the amine groups were formed on the surface of PDMS by silanization [25]. Then the PDMS was capped and brought into contact with SU-8 channels and by applying the pressure gently and keeping the integrated device for 2 hours in oven at temperature of 80°C, the bonding was carried out. Figure 2(a-d) shows the complete procedure for fabricating SU-8 based microfluidic channels and sealing these channels with PDMS. Figure 2(e) shows the image of on-chip device with one inlet and one outlet and Figure 2(f) shows the fabricated SU-8 channel. The idea of fabricating SU-8 based microfluidic channel and capping with PDMS, instead of fabricating conventional PDMS based microfluidic channel was to get rid of the precise alignment of the PDMS based microfluidic channels and the plasmonic nanostructures which is a very big challenge. Moreover, the bonding between plasma treated PDMS and gold surface was found to be very weak which results in leakage from the channels.

To test the bonding quality of these SU-8 based microfluidic channels, DI water was pumped through these channels at different flow rates (from 0.5 µl/ minute to 30 µl/minute) using a syringe pump and captured microscope images confirmed the flow of water without any leakage through the channel.

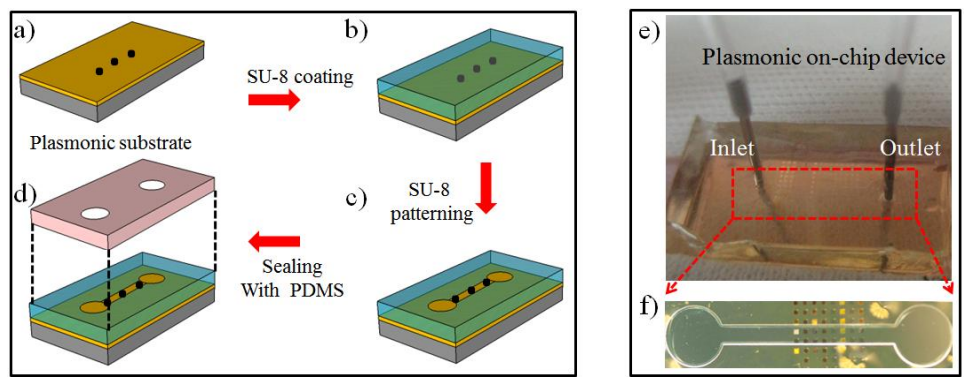

Fig. 2(a-d) : The process flow for fabrication of SU-8 based microfluidic channel (e) On-chip device with one inlet and one outlet (f) Microscopic image of fabricated SU-8 channel

**Characterization setup:**

Figure 3 shows the characterization setup used for real-time monitoring of refractive index induced surface modifications for different analytes. A pair of cross polarizer-analyzer ($\theta_P$ = 45° and $\theta_A$ = 135°) in an inverted broadband leakage radiation bright microscope is used to diminish direct $0^{th}$ order transmission and capture bright SPs emission (due to ±1 diffraction order) against a dark background. A polarization rotation of 90° in fabricated nanostructures mediated by SP excitation was used to develop a dark field plasmon polarization microscope using a conventional bright field microscope under two crossed axis polarizers [22]. Leakage of plasmon coupled transmitted radiation from the bottom of the substrate was collected by a 100x immersion oil objective lens. A flip mirror was used to divert the beam to a Fourier lens, and camera placed at its focus was used to obtain the FP images. Inset (i) and (ii) shows the FP and real plane images captured for 2D plasmonic nanostructures of period $P_x = P_y =$ 500 nm respectively. In real and FP images, the yellow brown color obtained due to the (0, ±1) and (±1, 0) diffraction order coupled to SP mode at cross polarizer-analyzer position. The detailed mechanism for this phenomenon has been described in our recent work [22].

**Capturing refractive index induced surface modifications:**

The analytes were infused into the device using 2 ml, 1000 series gas tight syringes actuated by syringe pump to control the flow rate of the analytes passing through microfluidic channel. The experiments were carried out inside an air-conditioned lab at a steady temperature of 25 °C to overcome the effect of temperature on the viscosity of different analytes [26]. The broadband white light source in an inverted microscope passing through a polarizer was incident on the device and the transmitted light was captured by real and FP CCD camera after passing through a cross axis analyzer.

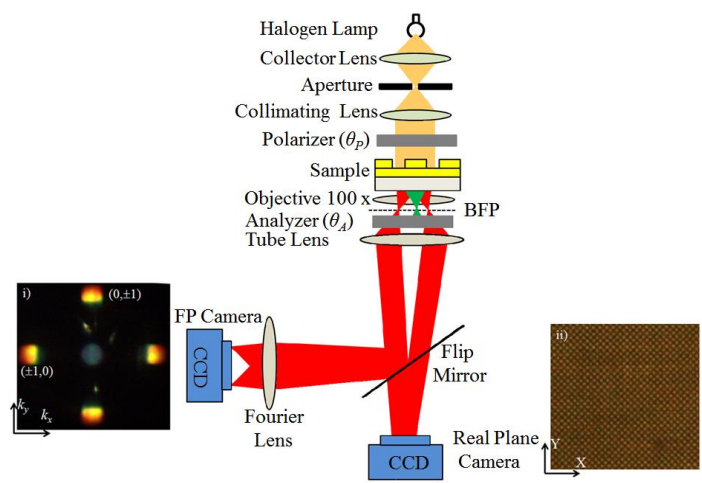

Fig. 3(a) : Characterization set up. Inset (i) and (ii) shows the captured FP and real plane images for 2D plasmonic nanostructures of $P_x = P_y = 500$nm

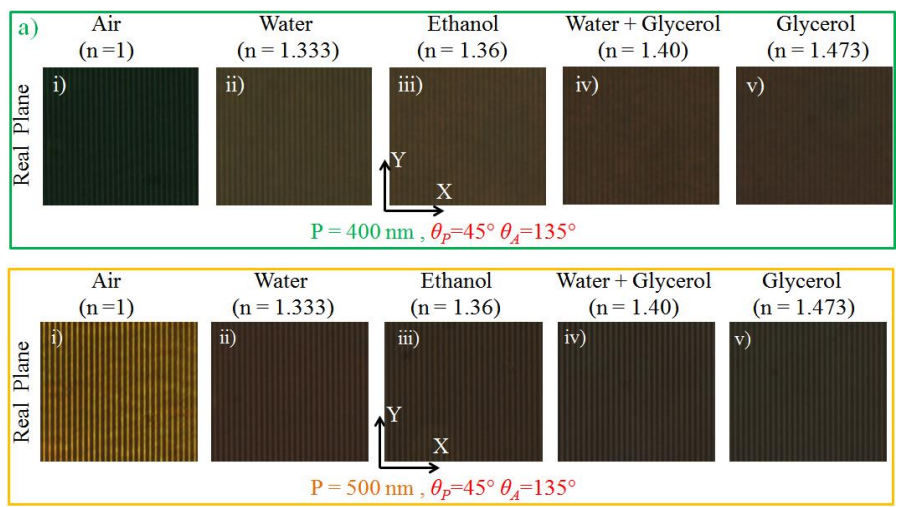

Fig. 4: Real plane images of 1D plasmonic nanostructure with (a) P = 400 nm (b) P = 500 nm with analyte of different refractive indices

The analytes of different refractive indices were introduced on the sensing sites through microfluidic channel having nanostructures with different periods. First the measurement was carried out with air and then the analyte with known refractive index was introduced.

Finally the channel was washed with DI water and exposed to air to make sure that analyte was completely washed off the surface and the color returns to that air reference before introducing the next analyte.

Figure 4(a) shows the real plane images for 1D plasmonic nanostructure of P = 400 nm with analytes of different refractive indices. The shift in color from Green to Orange-Red was clearly observed with increase in refractive index in captured images. In the same channel, the images for different analytes (as introduced in Figure 4(a)) for 1D plasmonic nanostructures of P = 500 nm were also captured as shown in Figure 4(b). The change in color with increase in refractive index can be clearly seen with analyte of different refractive indices with naked eyes. The change of color in captured images due to refractive index induced surface modification was quantified with CMYK components of images extracted using image processing [17,18] as shown in Figure 5. The shift in color corresponding to the SP resonance wavelength was observed with increase in the refractive index and an index resolution of $1.63 \times 10^{-4}$ RIU was experimentally achieved by extracting the change in average values of C, M, Y, K components from captured images.

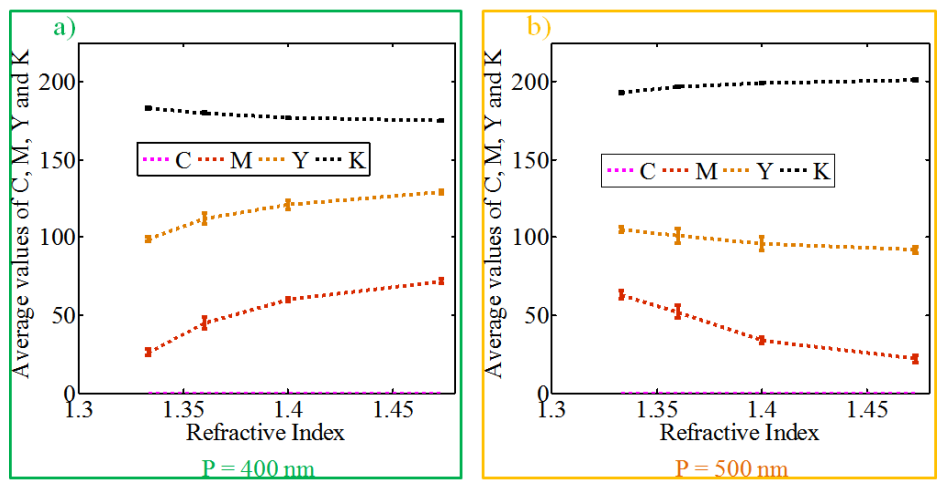

Fig. 5: Average values of C, M, Y and K for captured images with varying refractive index of analyte for 1D plasmonic structure of (a) P = 400 nm (b) P = 500 nm

**Label free imaging of miscible and immiscible analytes:**

The performance of on-chip plasmonic device with a Y shaped microfluidic channel integrated with plasmonic substrates was investigated by mixing of two colorless liquids. First, the analysis was started with two miscible liquids and then for immiscible liquids. Figure 6(a) shows the image of on-chip plasmonic device with Y shaped microfluidic channel and Figure 6(b) shows the Y shaped SU-8 microfluidic channel. The water and ethanol were taken as two miscible liquids to capture the interface of water and ethanol. The water was passed through the left inlet and the ethanol was passed through the right inlet. Figure 6(c) shows the mixing of water and ethanol without adding any colors where the interface of water and ethanol could not be distinguished.

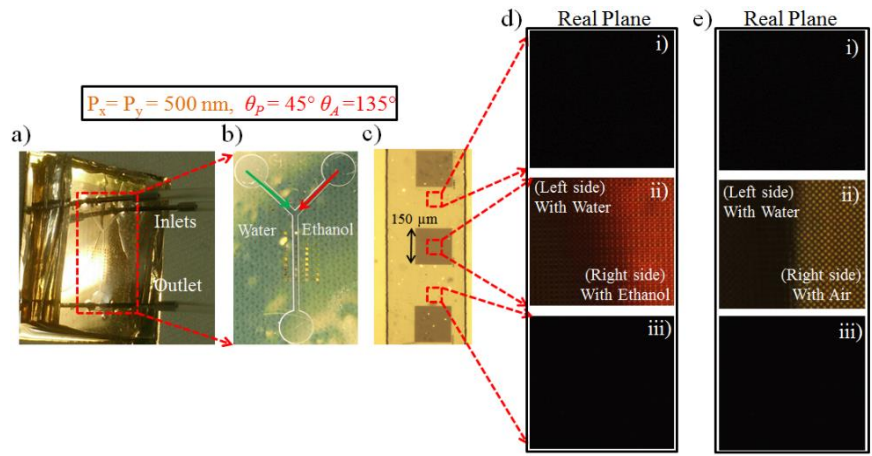

Fig. 6: (a) Image of on-chip plasmonic device with Y shaped channel (b) Microscopic image of Y shaped SU-8 microfluidic channel (c) Microscopic image for mixing of water and ethanol without adding any colors in microfluidic channel; Real plane images for without and with 2D plasmonic metallic pillars of $P_x = P_y = 500$ nm to capture
(d) Water - Ethanol interface (e) Water - Air interface

So here, the color-selective filtering nature of plasmonic nanostructures was used for observing the interface of two colorless miscible liquids while flowing in Y shaped channel. The real plane images were captured for the liquids flowing in the channel on the locations with and without plasmonic nanostructures. Figure 6-d (ii) shows the captured real plane image for 2D plasmonic nanostructure with metallic pillars of $P_x = P_y = 500$ nm. The difference in color for water and ethanol was clearly seen due to different refractive indices in captured real plane image, whereas, the images captured on the locations without plasmonic nanostructures were found completely dark as expected. An another experiment was also performed to compare our results with reference by changing the right inlet analyte (ethanol) to air and captured the water-air interface on 2D metallic pillars of $P_x = P_y = 500$ nm as shown in Figure 6-e(ii). Thus, the color-selective filtering nature of plasmonic nanostructures can be used for capturing the interface of two colorless miscible liquids without any stain (label).

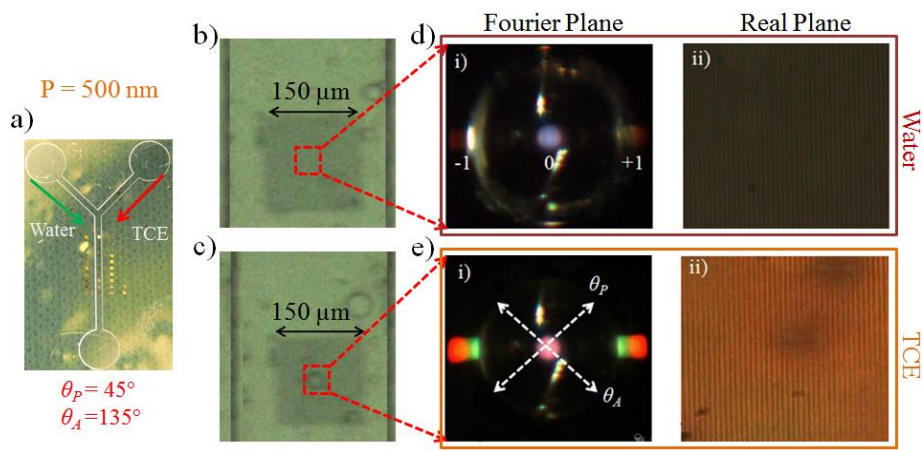

Fig. 7: (a) Image of on-chip plasmonic device with Y shaped channel (b) Microscopic images for mixing of two colorless immiscible liquids (b) Without TCE droplet (c)With TCE droplet; Fourier and Real plane images
for 1D plasmonic nanostructure of P = 500 nm with (d) Water (e) TCE

After capturing the interface of two colorless miscible liquids, the mixing of two immiscible liquids was studied. Here, the water and TCE ( Tri Chloro Ethylene) were taken as two immiscible liquids. While mixing of Water and TCE, TCE droplets were formed, which were captured in real and FP images for 1D plasmonic nanostructures of P = 500 nm. Figure 7(b) &

(c) show the microscopic images for mixing of two colorless immiscible liquids, without TCE droplet and with TCE droplet respectively. Figure 7(d) & (e) show the FP and real plane images for capturing water and TCE. Whenever TCE droplet passed through the sensing sites, the change in color was clearly observed in FP and real plane images as shown in Figure 7(c) & (d). In FP images, the change in color for ±1 diffraction order coupled to SP mode was observed from dark red to orange due to change in the index of water and TCE.

**Capturing air bubbles in Real and FP images using hydrodynamic focusing:**

Hydrodynamic focusing was also demonstrated on presented on-chip device where the width of the hydrodynamically focused stream of one liquid was controlled by the relative flow rates of the three liquids. The most common configuration is a 3-inlets-1-outlet device which was fabricated and tested that allowed to capture the flow of air bubbles on plasmonic nanostructures with real and FP images through hydrodynamic focusing.

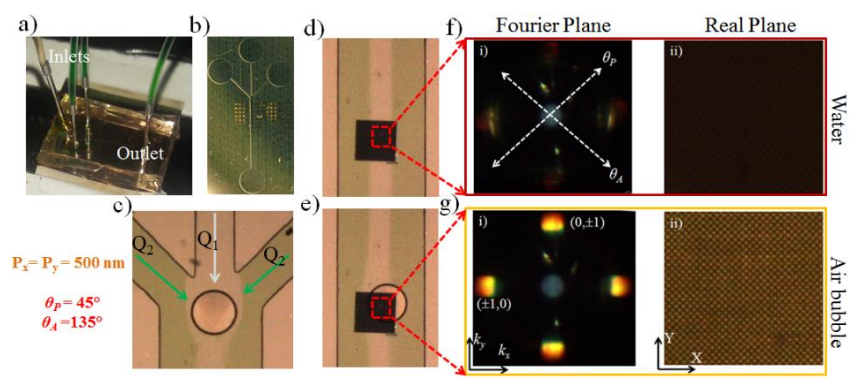

Fig. 8: (a) Image of on-chip device with 3 inlets and 1 outlet (b) Microscopic image of SU-8 channel with 3 inlets and 1 outlet (c) Flowing air bubbles while hydrodynamic focusing on the on-chip device; Microscopic images of the device with 3 inlets and 1 outlet (d) Without air bubble (e) With air bubble; FP and Real plane images of plasmonic nanostructures of $P_x = P_y = 500$ nm for capturing (f) Water (g) Air bubble

Figure 8(a-b) shows the image for on-chip device and SU-8 channel with 3 inlets and 1 outlet respectively. The DI water, with Green food color added manually (For visualization purposes), was passed through the outer inlets, whereas, the DI water without any color was passed through the middle inlet. The flow rate of the middle inlet was kept fixed to $Q_1 = 4$ µl/minute, while, the flow rate in left and right inlets was changed from 1.5 µl/minute to 4 µl/minute in step of 0.5 µl/minute with help of syringe pumps. The idea to change the flow rate of outer inlets from 1.5 µl/minute to 4 µl/minute in steps was to demonstrate the hydrodynamic focusing, in other words by varying the flow rates, it is possible to get the focused stream for middle inlet. After achieving the hydrodynamic focusing, the air bubbles were introduced through middle inlet with water as shown in Figure 8(c-e) and the images of the air bubbles passing through fabricated 2D nanostructures of $P_x = P_y = 500$ nm were captured in FP and real plane as shown in Figure 8(f) and (g) respectively. The change in color from dark Red to Yellow brown was clearly observed in FP and real plane images as air bubble passed through sensing sites (In this case 2D plasmonic nanostructures with metallic pillars of $P_x = P_y = 500$ nm). This imaging technique on proposed fabricated plasmonic structures can be used to count the number of air bubbles or cells for bio applications.

**Conclusions:**

The optofluidic integration using microfluidic channels on engineered fabricated plasmonic nanostructures was carried out to demonstrate real-time index sensing. The analytes of different refractive indices were introduced on the sensing sites through SU-8 microfluidic channels and the change in the color corresponding to SP excitation wavelengths was captured in real and FP images. An index resolution of $1.63 \times 10^{-4}$ RIU was achieved by quantifying CMYK components of images. Label-free imaging of the interface of colorless miscible and immiscible liquids was presented flowing in laminar flow regime in the microfluidic channels by using the color-selective filtering nature of plasmonic nanostructures. Hydrodynamic focusing was demonstrated on presented on-chip plasmonic device and the flow of air bubbles passing though channel on plasmonic nanostructures was captured in FP and real images due to change in refractive index from water to air. The purpose of the work is to establish the foundation of on-chip label-free imaging microscopy to avoid fluorescence tagging and to provide a simple platform where the optical image could be visually interpreted for surface modifications.

**Acknowledgment:**

The authors would like to thank Centre for NEMS and Nano Photonics (CNNP) for the use of fabrication facilities. The authors are thankful to Prof. Ashis Sen for allowing to use the facilities of Microfluidics lab and Dr. P. Sajeesh for many fruitful discussions related to microfluidics.